\documentclass[prb,twocolumn,showpacs,superscriptaddress,
amsmath,amssymb]{revtex4}
\usepackage{graphicx} 
\usepackage{amsmath} 
\usepackage{color}

\long\def\comment#1{}
\begin{document}

\title{Topological excitations in three dimensional Kitaev model}

\author{Saptarshi Mandal} \affiliation{ International Institute of
  Physics, UFRN, 59078-400, Natal-RN, Brazil}
\email{mandal.saptarshi1@gmail.com} \author{Naveen Surendran}
\affiliation{The Abdus Salam International Centre for Theoretical
  Physics, Strada Costiera 11, 34151 Trieste, Italy}
\altaffiliation{Present address: Max-Planck Institute for the Phyiscs
  of Complex Systems, Noethnitzer Strasse 38, D-01187 Dresden,
  Germany}\email{naveen@pks.mpg.de} \date{\today}

\begin{abstract}
  We study the excitations in a three dimensional version of Kitaev's
  spin-$\frac{1}{2}$ model on the honeycomb lattice introduced by the
  present authors recently. The gapped phase of the system is analyzed
  using a low energy effective Hamiltonian $H_{eff}$ which is defined
  on the diamond lattice and consists of plaquette operators. The
  excitations of $H_{eff}$ form loops in an embedded lattice. The
  elementary excitations, which are the shortest loops, are
  fermions.  Moreover, the excitations obey nontrivial braiding rules:
  when a fermion winds through a loop, the wave function acquires a
  phase $\pi$.
\end{abstract}

\pacs{75.10.Jm,03.67.Pp,71.10.Pm}
\maketitle

\section{\label{S-Int} Introduction}

Kitaev's spin-$\frac{1}{2}$ model on the honeycomb lattice has become
a paradigmatic system in the study of topological order in condensed
matter systems \cite{Kit06}. The model is exactly solvable; there is a
gapped phase that supports Abelian anyons and a gapless phase in which
the excitations (in the presence of a perturbing and gap inducing
external magnetic field) are non-Abelian anyons. A remarkable feature
of Kitaev's Hamiltonian is that, in contrast to other topologically
ordered solvable models such as the toric-code \cite{Kit03}, it
involves only local two-spin interactions, which makes it a physically
feasible model.
Various aspects of the model have been extensively studied so far
\cite{DuaDemLuk03,BasMan07,HanNus08,KaiDus08,DusKai08,BasSenSha08,SenSenMon08}.

The key element in Kitaev's construction, leading to its exact
solvability, is that the lattice is trivalent and that the links can
be labeled with three different colors in such a way that at each
site no two links have the same color. Using this fact, the present
authors have generalized his construction to three dimensions
\cite{ManSur09}. There exists other three dimensional
generalizations of the Kitaev model \cite{SiYu08, Ryu09}.

The 3D Kitaev model introduced in Ref. \onlinecite{ManSur09} has also
been solved exactly and has a phase diagram similar to the 2D case,
with a gapped phase and a gapless one. It has further been shown that
the gapped phase is described by an effective toric-code-like
Hamiltonian $H_{eff}$ defined on the diamond lattice and consisting of
plaquette operators. In this paper, we analyze the excitations of
$H_{eff}$. We find that they have the structure of loops, the energy
of a loop being proportional to its length.

The elementary excitations, which are along the smallest loops, are
fermions. Since the excitations are one dimensional objects, it is
meaningful to ask, even though they live in three dimensions, whether
they obey nontrivial braiding statistics. We find that the state
acquires a phase $\pi$ when a fermion winds through a loop.

The paper is organized as follows. Section \ref{S-Ham} briefly reviews
the three dimensional Kitaev model and the low energy effective
Hamiltonian $H_{eff}$ describing its gapped phase. In Sec. \ref{S-Exc}
we give the representation of the excitations of $H_{eff}$ as loops in
an embedded lattice. Calculation of exchange and braiding statistics
is presented in Sec. \ref{S-Sta}. We conclude with a discussion of our
results in Sec. \ref{S-Dis}.

\section{\label{S-Ham} Hamiltonian}
\begin{center}
\begin{figure}[ht]
\includegraphics[width=.42\textwidth]{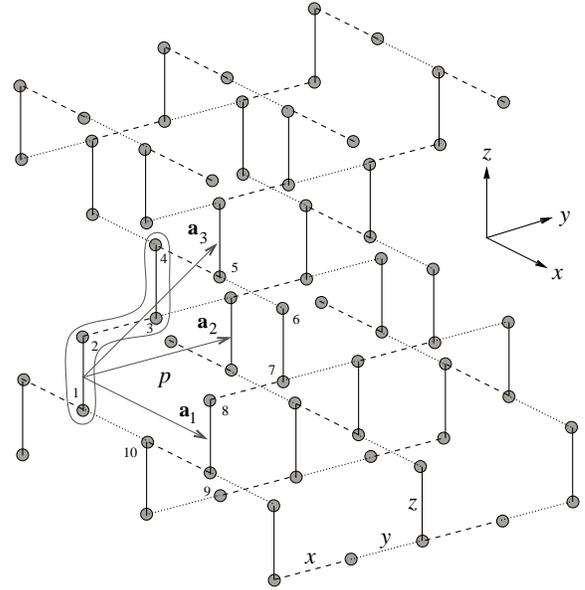}
\caption{ \label{fig-3dlattice} The 3D lattice: The four sites inside
  the loop (marked 1 to 4) constitute a unit cell; ${\bf a}_1, {\bf
    a}_2$ and ${\bf a}_3$ are the basis vectors. Plaquette
  $p$ consists of sites marked 1 to 10.}
\end{figure}
\end{center}

The Hamiltonian is defined on the lattice shown in Fig. \ref{fig-3dlattice}.
For details about the lattice we refer to
Ref. \onlinecite{ManSur09}. Here we just note that the
coordination number is three and the links, which connect
neighboring sites, are labeled $x$, $y$ or $z$. In addition, the
three links at each site have different labels.
The Hamiltonian is
\begin{equation}
\label{E-KitHam}
H = -J_x \sum_{<i,j>_x} \sigma_i^x \sigma_j^x - J_y \sum_{<i,j>_y} \sigma_i^y
\sigma_j^y -J_z \sum_{<i,j>_z} \sigma_i^z \sigma_j^z,
\end{equation}
where $\sigma^a_i$ are the Pauli matrices at site $i$ and $\sum_{<i,j>_a}$
denotes summation over $a$-type links.

The above Hamiltonian can be solved exactly\cite{ManSur09}
using a Majorana fermion representation of Pauli
matrices\cite{Kit06}. The excitation spectrum is gapless when \mbox{$J_z
  \le J_x + J_y$}, \mbox{$J_x \le J_y + J_z$}, and \mbox{$J_y \le J_z
  + J_x$}. Everywhere else in the parameter space there is a gap. The
purpose of this paper is to study the nature of low-energy excitations in the
gapped phase. Specifically, we will concentrate on the limit $J_z \gg J_x,
J_y$.

Let $H_0$ be the Hamiltonian obtained by putting \mbox{$J_x=J_y=0$} in
Eq. (\ref{E-KitHam}); it corresponds to isolated $z$-links and the
spectrum is trivially solved. The ground state has a large degeneracy:
any state in which the two spins in any given $z$-link, say, connecting
sites $i$ and $j$, are either \mbox{$|\uparrow\uparrow\rangle_{ij}$}
or \mbox{$|\downarrow\downarrow\rangle_{ij}$} has the minimum
energy. Here \mbox{$\sigma_{i}^z |\uparrow\rangle_{i} =
  |\uparrow\rangle_{i}$}, \mbox{$\sigma_{i}^z |\downarrow\rangle_{i} =
  - |\downarrow\rangle_{i}$}, and similarly for $j$. 

\begin{figure}[ht]
\begin{center}
\includegraphics[width= .27\textwidth]{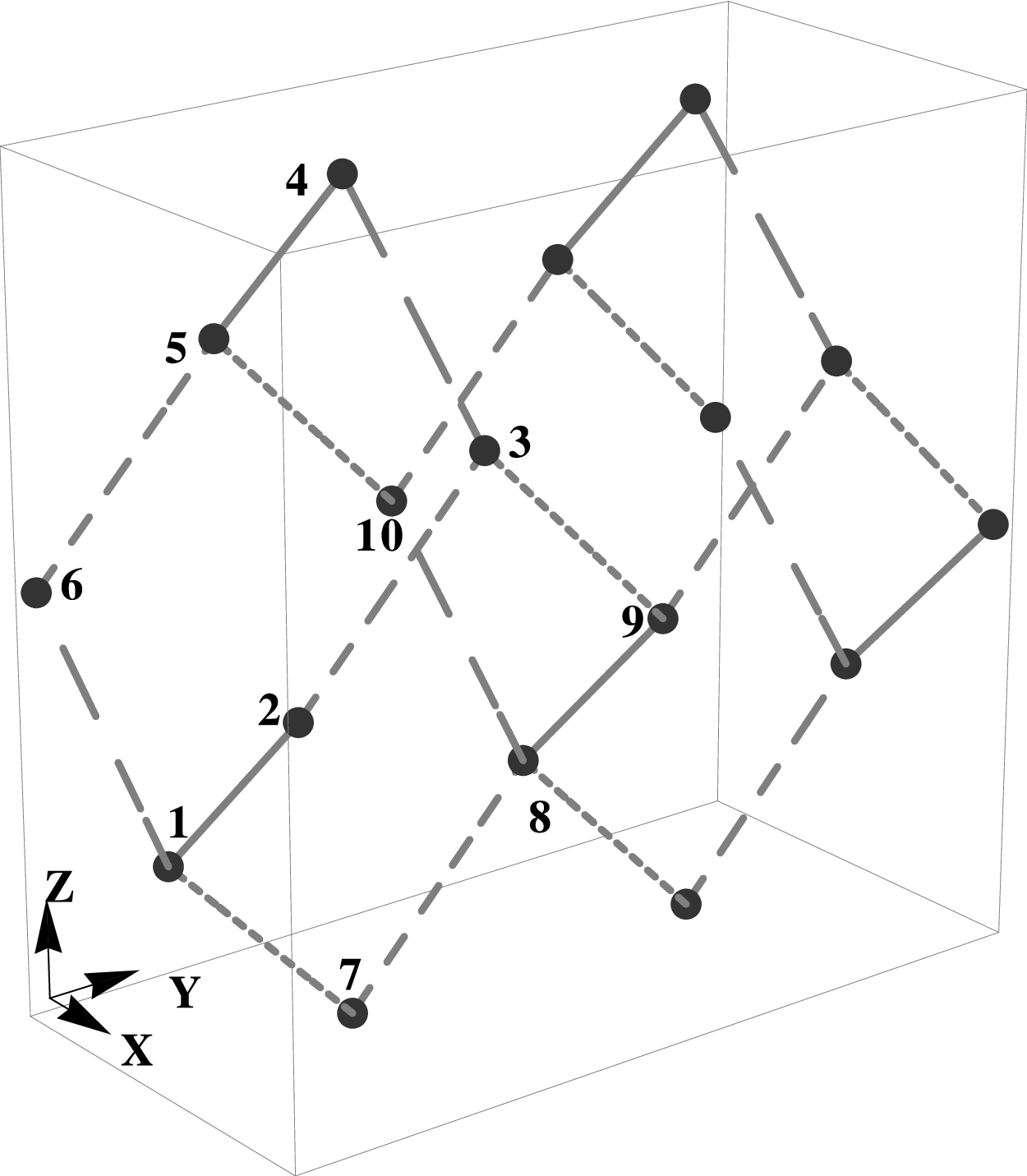}
\caption{ \label{F-EffLat} The lattice $\mathcal{L}_{eff}$ (diamond
  lattice) on which the effective Hamiltonian is defined. Bold,
  long-dashed, medium-dashed and short-dashed lines respectively
  denote links with labels $a, b, c$ and $d$.}
\end{center}
\end{figure}

For small values of $J_x$ and $J_y$, the low energy excitations (more
precisely, those eigenstates of $H$ adiabatically evolving from the
degenerate ground states of $H_0$, as $J_x$ and $J_y$ are turned on)
are described by an effective Hamiltonian $H_{eff}$ acting on this
degenerate subspace (denoted by $\mathcal{H}_0$).  Such an effective
Hamiltonian will be defined on the lattice formed by the $z$-links,
which we call $\mathcal{L}_{eff}$, and it turns out to be the diamond
lattice, as far as the connectivity is concerned (see
Fig. \ref{F-EffLat}). $\mathcal{L}_{eff}$, which has coordination
number four, is obtained from the original lattice by shrinking each
$z$-link to its mid-point. 


The links in $\mathcal{L}_{eff}$ lie along four different
directions. We can choose an orthogonal coordinate system such that
these directions are $(1,0,1),~(-1,0,1),~(0,-1,-1)$ and $(0,1,-1)$. We
label the links respectively as $a, b, c$ and $d$. (Note that the
links are not directed; positive and negative directions are
equivalent.) Then at each site, the four links have four different
labels.

Let $I$ denote the sites of $L_{eff}$. $\mathcal{H}_0$ is spanned by
the states \mbox{$\big\{\prod_I |m_I\rangle_I ~|~ m_I \in
  \{+,-\}\big\}$}, where \mbox{$|+\rangle_I \equiv
  |\uparrow\uparrow\rangle_I $} and \mbox{$|-\rangle_I \equiv
  |\downarrow\downarrow\rangle_I $}.  The local degrees of freedom
acting on $\mathcal{H}_{0}$ are effective \mbox{spin-$\frac{1}{2}$}
operators $\tau^\alpha_I$ (\mbox{$\alpha \in \{x,y,z\}$}) defined by
\begin{eqnarray}
\label{E-TauDef}
\tau_I^z |+\rangle_{I} = |+\rangle_{I}, && \tau_I^z |-\rangle_{I} = 
 - |- \rangle_{I}, \nonumber \\
\tau_I^x |+ \rangle_{I} = |-\rangle_{I}, && \tau_I^x |- \rangle_{I} = 
 - |+ \rangle_{I}.
\end{eqnarray}

Before writing down $H_{eff}$, it is convenient to define a loop
operator as follows. Consider a loop $l$ in $\mathcal{L}_{eff}$ of
length $n$ and consisting of sites \mbox{$\{I_k | k=1,2, \dots n\}$}
such that $I_k$ and $I_{k+1}$ are nearest neighbors, with the
identification $I_{n+1} = I_1$. Let $\gamma_k$ denote
the link  connecting sites $I_k$ and
$I_{k+1}$. We then define
\begin{equation}
  \label{E-LooOpe}
B_l = \prod_{k=1}^{n} \tau_{I_k}^{\alpha_k},
\end{equation}
where $\alpha_k$ is determined by the labels of the links $\gamma_{k-1}$
and $\gamma_k$ as follows: 
\begin{eqnarray}
\label{E-Rul}
\alpha_k = x && \mbox{for labels $(a,c)$ and $(b,d)$}, \nonumber \\
\alpha_k = y && \mbox{for $(a,d)$ and $(b,c)$},\nonumber\\
\alpha_k = z && \mbox{for $(a,b)$ and $(c,d)$}. 
\end{eqnarray}
Here we need to consider only pairs with different labels since
$\gamma_{k-1}$ and $\gamma_k$ are both links involving site $k$ and
all links at a given site have different labels. It is easy to
see that, with this definition, for any two loops $l$ and $l^\prime$,
\begin{equation}
\label{E-LooCom}
[B_l, B_{l^\prime}]=0.
\end{equation}
This can be understood as follows. First, note from Eqs. (\ref{E-Rul})
that $\alpha_k$ is the same for two pairs of labels if and only if
they are complementary to each other. Now, $l$ and $l^\prime$ can
intersect each other at one or more sites, and they can also have
overlapping segments. Let $I$ be a site where they intersect. Then the
two pairs of labels for the links that meet at $I$, corresponding to
$l$ and $l^\prime$, will be complementary to each other and therefore
$\alpha_k$ will be the same for both the loops.  Now consider a
segment where they overlap. Again, $\alpha_k$ will be the same for $l$
and $l^\prime$ everywhere on the segment except at the two end points,
where they are strictly different. This is true for any overlapping
segment of $l$ and $l^\prime$, which in turn means that among those
sites shared by the two loops, there will be an even number of them
where $\alpha_k$ is different for the two loops. This implies
Eq. (\ref{E-LooCom}) since $\tau_I^a \tau_I^b = -\tau_I^b \tau_I^a$
when $a\ne b$.

To lowest order in $J_x$ and $J_y$, $H_{eff}$ is given in terms of the
above operators corresponding to the smallest loops in
$\mathcal{L}_{eff}$. We call such loops plaquettes and use the index
$p$ to denote them. A plaquette consists of six sites and there are
four types, each type corresponding to a particular label sequence and
orientation. The four possible sequences are $acbacb$ (type-1),
$cadcad$ (type-2), $bdabda$ (type-3) and $dbcdbc$ (type-4). In
Fig. \ref{F-EffLat}, $\{1,2,3,4,5,6\}$,
$\{7,8,9,3,2,1\}$, $\{8,10,5,4,3,9\}$ and $\{7,1,6,5,10,8\}$ are
respective examples. 

In terms of the plaquette operators, the lowest order effective
Hamiltonian is \cite{ManSur09}
\begin{equation}
H_{eff} =  -\frac{7}{256J_z^5} \left( J_x^4 J_y^2 \sum_{p}  B_{p}
+ J_x^2 J_y^4 {\sum_{p}}^\prime   B_{p} \right), 
\label{E-EffHam}     
\end{equation}
where the unprimed sum is over plaquettes of type- $1$ and $2$, while
the primed sum is over plaquettes of type- $3$ and $4$.

\section{\label{S-Exc} Excitations}

Since all plaquette operators commute among themselves [see
Eq. (\ref{E-LooCom})], $H_{eff}$ is, in principle, trivially
diagonalized. The energy eigenstates are then labeled by the
eigenvalues of $B_p$, which are $+1$ or $-1$. However, not all $B_p$
are independent. In fact, it is the constraints existing among them
that make $H_{eff}$ nontrivial and give rise to topological
excitations. 

Next we write down all the constraints and then find the allowed
configurations of $B_p$ that are consistent with the constraints. In
the process we will also calculate the ground state degeneracy. From
now on we assume periodic boundary conditions in all three directions.

We first find a complete set of mutually commuting conserved operators
that includes a maximal number of plaquette operators. To this end, we
define a product of two loops. For our purposes, it is convenient to
represent a loop by the set of links it contains. Let $l_1 =
\{\gamma_1, \gamma_2, \dots \gamma_{n_1}\}$ and $l_2 = \{\delta_1,
\delta_2, \dots \delta_{n_2}\}$ be two loops in $\mathcal{L}_{eff}$
containing $n_1$ links (denoted $\gamma_i$) and $n_2$ links (denoted
$\delta_i$) respectively. The product of $l_1$ and $l_2$ is then
defined as
\begin{equation}
l_{12} \equiv l_1 \cdot l_2= l_1 \cup l_2 - l_1 \cap l_2.
\label{E-LooPro}
\end{equation}
That is, $l_{12}$ is the loop formed by the links contained in $l_1$
or $l_2$ but without those links common to both.
Then it is easy to check that 
\begin{equation}
\label{E-LooCon}
B_{l_{12}} = B_{l_{1}} \cdot B_{l_2}.
\end{equation}
Therefore, to form a complete set of conserved operators we need to
consider only the elementary plaquettes, but not any loop that can be
obtained by combining a subset of the latter. There are three
topologically nontrivial (noncontractible) loops corresponding to the
three directions, which cannot be obtained from the plaquettes; we
denote them $C_x$, $C_y$ and $C_z$ respectively. Any loop in the
lattice can be obtained by combining the plaquettes and the three
noncontractible loops. However, as we stated earlier, the plaquettes
themselves are not all independent. There are three types of
constraints that plaquette operators satisfy.

\paragraph{Local constraints.} It follows from Eq. (\ref{E-LooCon})
that the product of any set of plaquettes that forms a closed surface
equals 1. The smallest such surface is formed by four adjacent
plaquettes, any two of which share two links, as shown in
Fig. \ref{F-EffLat}. We call the volume enclosed by such a surface a
\emph{basic cell}. There are $N$ basic cells, where $N$ is the number
of sites. This gives rise to $N-1$ independent constraints; it is
$N-1$ because the constraint corresponding to any one basic cell
equals the product of the constraints corresponding to the remaining
$N-1$.

\paragraph{Surface constraints.} As a consequence of
Eq. (\ref{E-LooCon}) there are three further constraints, independent
of the local constraints, which correspond to the three topologically
nontrivial surfaces.

\paragraph{Volume constraints.} Any product of corner sharing
plaquettes that fill out the whole lattice equals one. There is only
one such independent constraint. 

Putting everything together, the total number of constraints is
$N+3$. There are $2N$ plaquettes and the number of independent
plaquette operators is $N-3$. These along with the three 
noncontractible loop operators $C_x$, $C_y$
and $C_z$ form a complete set of $N$ commuting operators.

There is another set of conserved operators corresponding to
noncontractible surfaces. Such an operator can be defined on a plane
$S$ that divides the lattice without cutting any links in such a way
that at each site lying on the plane, two of the links lie on one side of
the plane and the remaining two on the other. Then the surface
operator is defined as follows:
\begin{equation}
\label{E-SurOpe}
M_S =  \prod_{I\in S} \tau_I^{\alpha_I},
\end{equation}
where the product is over sites lying on $S$ and $\alpha_I$ is
determined by the two incoming (or outgoing) links according to the
rule given in Eq. (\ref{E-Rul}). $M_S$ commutes with all
contractible loop operators because noncommuting---in our case,
anticommuting---contributions come only at points where the loop
pierces the surface and there will be an even number of such points.
There are three such operators corresponding to the planes normal to 
 $x$, $y$ and $z$ direction. We denote them $M_X$, $M_y$
and $M_z$, respectively.

Just as we obtained loop operators as a product of link-sharing
plaquette operators, we can construct conserved `surface' operators by
multiplying corner sharing plaquette operators. In such a product,
contribution from a shared site cancel out since it will be the same
Pauli matrix for the two plaquettes sharing it [owing to
  Eq. (\ref{E-Rul})].  As a result the product will only involve sites
that are not shared. Consequently, when corner sharing plaquettes fill
out a certain volume in the lattice, the contribution to the operator
will come only from the sites on the boundary.

The set of noncontractible loop and surface operators $C_x$, $C_y$,
$C_z$ and $M_X$, $M_y$, $M_z$ form a system of three
spin-$\frac{1}{2}$ degrees of freedom since
\begin{eqnarray}
C_a M_a = -M_a C_a,&& C_aM_b = M_bC_a~\mbox{for}~ a\ne b, \nonumber \\
C_a C_b = C_b C_a,&& M_aM_b = M_bM_a.
\end{eqnarray}
It then follows that the ground state degeneracy is $2^3 =8$, since
the Hamiltonian does not depend on the above nonlocal operators. In
fact, all excitations will have the same degeneracy of 8.

In the ground state $|GS\rangle$, $B_p = +1$ for all
plaquettes. Evidently, all the constraints are satisfied. The
excitations are obtained by flipping some of the $B_p$'s to $-1$. To
find the configurations of $B_p$ that are consistent with the above
constraints, it is useful to consider the lattice formed by the basic
cells, each of which correspond to four plaquettes forming a closed
surface. This lattice, denoted $\mathcal{L}_p$, is also the diamond
lattice and the links in $\mathcal{L}_p$ represent the plaquettes in
the original lattice.

In  $\mathcal{L}_p$, the local constraints have a simple picture: the
plaquette operators corresponding to the four links at each vertex
multiply to 1. This implies that at every vertex only an even number
of links (0, 2 or 4) can be flipped to -1, which in turn means that
links with $B_p = -1$ form closed loops in  $\mathcal{L}_p$. It is
easy to see that any such loop excitation will also satisfy the surface
constraints. However, the volume constraint imposes a further
restriction on the allowed loop excitations. 

The smallest loop in $\mathcal{L}_p$ is an elementary plaquette
consisting of six links; we call it a 6-loop to distinguish it from
the plaquette in $\mathcal{L}_{eff}$. Nevertheless, the excitation of
an odd number of 6-loop violates the volume constraint. This can be
shown by explicit construction, but there is another more transparent
way to see this.

We have seen that the loop operators in Eq. (\ref{E-LooOpe}) leave the
ground state invariant. However, an open-string operator, obtained
from a loop operator by truncation, i.e. by removing the part
corresponding to a segment, is not conserved. The noncommuting part is
at the ends of the string; it anticommutes with 6 of the $B_p$'s
involving the spin at a given end. The plaquettes corresponding to
these 6 $B_p$'s form a 6-loop in $\mathcal{L}_p$. An open-string
operator then creates two 6-loop excitations at the ends. If two
excited loops $l_1$ and $l_2$ overlap, then the resultant excitation
will be along the loop $l_{12}$, which is the product of $l_1$ and
$l_2$ [see Eq. (\ref{E-LooPro})]. This is because the overlapping
links are flipped twice---equivalent to no flip.  Since any
noncontractible loop $l$ can be expressed as a product of 6-loops, the
excitation along $l$ can be created by exciting the corresponding
6-loops. 

In the extreme case in which the string consists only of one site, say
$I$, the corresponding string operator is $\tau_I^x$, $\tau_I^y$ or
$\tau_I^z$. Then the two 6-loops at the `ends' overlap (two of the
links are shared), and the resultant excitation is an 8-loop. Since
any state in the Hilbert space must be obtained from the ground state
by the action of a combination of $\tau_I^{\alpha}$ operators, it
follows that all states will have an even number of 6-loops, the
elementary excitations.

However, a single 6-loop can be isolated by taking the other end of
the string that creates it far away. This does not cost energy since
the loops do not interact except when they overlap (then the overlapping
part is annihilated). The energy of a loop is proportional to
its length, which corresponds to the number of $B_p$'s excited. 

In other words, the excitation spectrum has two superselection
sectors: 1) the vacuum, which consists of states with an even number
of 6-loop excitations and 2) the 6-loop, which consists of states with
an odd number of 6-loop excitations.

In general, an excitation corresponds to a configuration of one or
more loops in $\mathcal{L}_p$, where the loops can intersect but not
overlap. As we have just shown, such an excitation is created by
exciting the 6-loops into which the loop configuration can be
decomposed. Let $L$ be a loop configuration and let $\Gamma$ be one of
the decompositions of $L$ into 6-loops. It is straightforward to
explicitly construct the operator that creates the excitation along
$L$: arbitrarily pair up the constituent 6-loops; connect each pair by
a string, again arbitrarily; then take the product of all the string
operators that excite each pair.
  
Here it should be stressed that only the loop configuration $L$, on
which the excitation lives, is physical. The state depends neither on
the particular choice of decomposition $\Gamma$ nor on the specific
way in which the 6-loops are paired and connected by strings. All
choices give rise to the same state, up to a phase.

A loop excitation can also be interpreted as the edge of a membrane.
Just as open-string operators are obtained by truncating a loop
operator, we can construct an open-membrane operator by removing a
patch from a closed-surface operator defined as a product of corner
sharing plaquettes. Let $O_M$ be such an operator defined on a
membrane $M$. As in the case of string operators, $O_M$ does not
commute with the Hamiltonian. Nonetheless, it is only certain $B_p$'s
along the edge of $M$ that do not commute with $O_M$; in fact, they
anticommute. Therefore, in the state $O_M |GS\rangle$ these
anticommuting $B_p$'s are excited; they will correspond to a loop in
$\mathcal{L}_P$.  In the language of the previous construction, $O_M$
corresponds to a specific choice of decomposition into 6-loops, which
fixes $M$, and then neighboring 6-loops are paired together and
connected by the shortest string between them, which consists of a
single site.

\section{\label{S-Sta} Statistics}

In this section, we determine the statistics obeyed by the
excitations. We will see that the elementary 6-loop excitations are
fermionic. Nontrivial phases also arise from braiding: when a 6-loop
winds through a bigger loop, the state acquires a phase $\pi$.

\subsection{Localized excitations}
A 6-loop excitation is the maximally localized state permitted by the
constraints. So, they can be treated as particle-like and then one can
ask whether they are fermions or bosons. We recall that in three
dimensions there is no other statistics for point particles.

In a lattice, the statistics is determined by the algebra of the
hopping operators \cite{LevWen03}. Let $i,j,k,l$ be four sites in
the lattice and $\hat{t}_{\alpha \beta},~\alpha, \beta
\in\{i,j,k,l\},$ be the corresponding hopping operators. Then,
\begin{eqnarray}
\label{E-HopAlg}
\hat{t}_{lk} \hat{t}_{il} \hat{t}_{lj} &=&
\hat{t}_{lj}\hat{t}_{il}\hat{t}_{lk} ~ \mbox{for bosons}, \nonumber \\
\hat{t}_{lk} \hat{t}_{il} \hat{t}_{lj} &=&
-\hat{t}_{lj}\hat{t}_{il}\hat{t}_{lk} ~ \mbox{for fermions}. 
\end{eqnarray}

A 6-loop in $\mathcal{L}_p$ represents 6 plaquettes in
$\mathcal{L}_{eff}$, the lattice on which $H_{eff}$ is defined. In
$\mathcal{L}_{eff}$, such a set of 6 plaquettes share a common
link. Moreover, there is no other plaquette of which this particular
link is part of. This provides a unique one-to-one mapping
between links in $\mathcal{L}_{eff}$ and 6-loops in
$\mathcal{L}_p$. Therefore 6-loop excitations can be thought of as
localized around the links in $\mathcal{L}_{eff}$. 

The above mapping gives a description of loops in $\mathcal{L}_p$ in
terms of a link variable in $\mathcal{L}_{eff}$: it takes value $-1$
if the corresponding 6-loop is excited, otherwise, $+1$. This
description, however, is redundant due to a $\mathbb{Z}_2$ gauge
symmetry related to the non-uniqueness of the decomposition of a loop
into 6-loops. The gauge transformation at a given site flips all the
link variables connected to that site.

\begin{figure}[ht]
\begin{center}
\includegraphics[width= .1\textwidth]{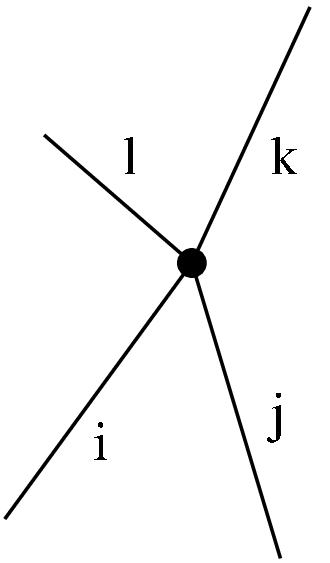}
\caption{ \label{F-FouLin} A site and the four links connected to it
  in the lattice $\mathcal{L}_{eff}$}
\end{center}
\end{figure}

Consider a site $I$ in $\mathcal{L}_{eff}$ and the four links
emanating from it, denoted $i,j,k,l$ (see Fig. \ref{F-FouLin}). In the
previous section, we showed that $\tau_I^{\alpha}$ excites an 8-loop,
which can be decomposed into two 6-loops. By the above mapping, this
corresponds to a pair of links at $I$, the value of $\alpha$
determining which pair. Note that, due to the gauge symmetry, exciting
a given pair or its complement are equivalent.

If $\tau_I^\alpha$ excites the links, say, $i$ and $j$, then it also hops an
excitation between the two links. Without any loss of generality, we
can assume that the links are labeled such that the hopping operators
are
\begin{eqnarray}
\hat{t}_{ij} = \hat{t}_{kl} = \tau_I^{x}, && \nonumber \\
\hat{t}_{ik} = \hat{t}_{jl} = \tau_I^{y}, && \nonumber \\
\hat{t}_{il} = \hat{t}_{jk} = \tau_I^{z}. && 
\label{E-HopOpe}
\end{eqnarray}
The first equality in each of the above equations follows from the
gauge symmetry. Moreover, $\hat{t}_{ij} = \hat{t}_{ji}$. It
immediately follows that
\begin{equation}
\label{E-HopAlg1}
\hat{t}_{lk} \hat{t}_{il} \hat{t}_{lj} =
-\hat{t}_{lj}\hat{t}_{il}\hat{t}_{lk}.
\end{equation}
Therefore, 6-loops are fermions.

\subsection{Braiding} 

A configuration of loops can evolve along topologically distinct
paths. The question one can ask then is whether there is a phase
difference between distinct paths. We will now show that a phase $\pi$
is acquired when a 6-loop winds through a bigger loop.

Let $O_M$ be a membrane operator that creates an excitation along loop $l$.
\begin{equation}
O_M = \prod_{I\in M} \tau_I^{\alpha_I}.
\end{equation}
Here $M$ is a set of sites $I$ which fills the surface enclosing
$l$, and $\alpha_I$ is determined by the labels of the two
incoming (or outgoing) links, according to Eq. (\ref{E-Rul}). 

\begin{figure}[ht]
\begin{center}
\includegraphics[width= .3\textwidth]{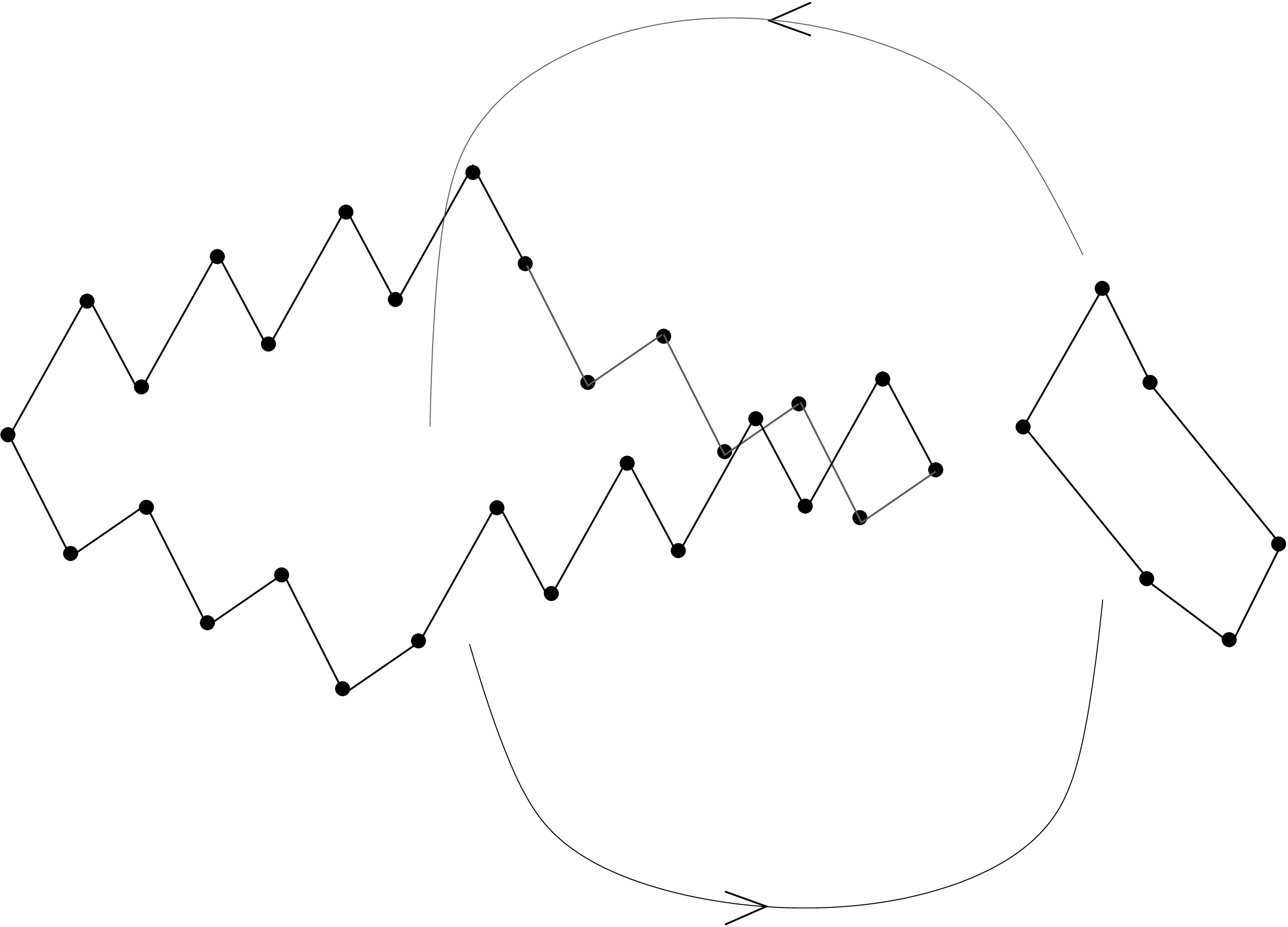}
\caption{ \label{F-Bra} Schematic depiction of a fermionic 6-loop
  braiding through a bigger loop. Traversing such a path results in a
  topological phase $\pi$}
\end{center}
\end{figure}

A 6-loop is moved along a closed path by the loop operator
corresponding to the path. A path which takes it through $l$ will have
one site common with $M$. Here we take $l$ to be big enough so that
the 6-loop can pass through without interacting, i.e., without
touching (see Fig. \ref{F-Bra}). Let this site be $I_c$. The loop operator
$B_l$ is given by
\begin{equation}
B_l = \prod_{I\in l} \tau_I^{\beta_I},
\end{equation} 
where $\beta_I$ are also determined by the set of rules in
Eq. (\ref{E-Rul}).

Since $l$ goes through the membrane, $\beta_{I_c}$ is determined by
one outgoing and one incoming link; therefore, it is strictly
different from $\alpha_{I_C}$. Hence
\begin{equation}
O_M B_l = - B_l O_M.
\end{equation}
This results in a phase difference of $\pi$ between paths which go
through $l$ and paths which do not.

\section{\label{S-Dis} Summary and  discussion}

We have studied the gapped phase of 3D Kitaev model by analyzing the
effective Hamiltonian $H_{eff}$ obtained in the limit of strong
$z-$links. $H_{eff}$ is defined on the diamond lattice and the
individual terms it contains are mutually commuting plaquette
operators.  There are loop and surface operators that are
conserved. In the language of Levin and Wen \cite{LevWen05}, our model
has both string-net and membrane-net condensation.

There is an embedded lattice $\mathcal{L}_p$, also a diamond lattice,
in which the links stand for plaquettes. In $\mathcal{L}_p$, the
excitations have the structure of loops. The energy of a loop is
proportional to its length, consequently, the ground state is free of
loops. The loops interact only when they overlap, in which case the
overlapping part is annihilated. The smallest loops in
$\mathcal{L}_p$, the 6-loops, are the elementary excitations and they
obey fermionic statistics. The fermionic 6-loop braids nontrivially
with a bigger loop: when it winds through the latter over a closed
path, the wave function acquires a phase $\pi$.

Even though there already exists other exactly solvable models in 3D
that are topologically ordered
\cite{LevWen03,HamZan05,BomMar07,Kim10}, our model has a particular
relevance in that it is obtained as the effective Hamiltonian for the
gapped phase of 3D Kitaev model, which involves only two-spin
interactions as opposed to higher-spin operators in the other solvable
models. It will therefore be worthwhile to analyze the model at finite
temperature to see if topological order survives
\cite{CasCha08,HamAli09,Kim10}.

\bibliography{kitaev}

\end{document}